\documentclass[12pt]{l4dc2020} 

\usepackage{hyperref}
\usepackage{graphicx}		
\usepackage{wrapfig}
\usepackage[format=plain,font=footnotesize,labelfont=bf,labelsep=period]{caption}
\usepackage{sidecap} 
\usepackage[export]{adjustbox}

\usepackage{amsmath}
\usepackage{amssymb}
\usepackage{mathtools}
\usepackage{algorithm}
\usepackage{algpseudocode}

\usepackage{pdfsync}
\usepackage[normalem]{ulem}
\usepackage{paralist}	
\usepackage[space]{grffile} 

\usepackage{color}
\newtheorem{assumption}{Assumption}

\newcommand{\R}{\mathbb{R}}

\definecolor{darkblue}{RGB}{0,0,102}
\definecolor{lightblue}{RGB}{77,77,148}

\definecolor{gold}{RGB}{234, 170, 0}
\definecolor{metallic_gold}{RGB}{139, 111, 78}

\renewcommand{\cal}[1]{\mathcal{ #1 }}
\newcommand{\mb}[1]{\mathbf{ #1 }}

\newcommand{\interior}{\mathrm{int}}

\DeclareMathOperator*{\argmin}{argmin}

\newcommand{\derp}[2]{\frac{\partial #1 }{\partial #2 }}

\usepackage{wrapfig}
\usepackage[format=plain,font=footnotesize,labelfont=bf,labelsep=period]{caption}
\usepackage{sidecap} 
\usepackage[export]{adjustbox}

\title[Learning Control Barrier Functions]{Learning for Safety-Critical Control with Control Barrier Functions}
\usepackage{times}

\author{%
 \Name{Andrew J. Taylor} \Email{ajtaylor@caltech.edu}\\
 \addr Caltech, Pasadena, CA, USA
 \AND
 \Name{Andrew Singletary} \Email{asinglet@caltech.edu}\\
 \addr Caltech, Pasadena, CA, USA
 \AND
 \Name{Yisong Yue} \Email{yyue@caltech.edu}\\
 \addr Caltech, Pasadena, CA, USA
 \AND
 \Name{Aaron D. Ames} \Email{ames@caltech.edu}\\
 \addr Caltech, Pasadena, CA, USA%
}

\begin{document}

\maketitle

\begin{abstract}%
 Modern nonlinear control theory seeks to endow systems with properties of stability and safety, and have been deployed successfully in multiple domains. Despite this success, model uncertainty remains a significant challenge in synthesizing safe controllers, leading to degradation in the properties provided by the controllers. This paper develops a machine learning framework utilizing Control Barrier Functions (CBFs) to reduce model uncertainty as it impact the safe behavior of a system. This approach iteratively collects data and updates a controller, ultimately achieving safe behavior. We validate this method in simulation and experimentally on a Segway platform.
\end{abstract}


\section{Introduction}
Safety is of significant importance in many modern control applications, and as the complexity of systems increases, it is necessary to rigorously encode safety properties during the controller design process. Autonomous driving, industrial robotics, and aerospace vehicles are examples of rapidly growing applications where safe control synthesis is critical. In practice, the models used to design such controllers are imperfect, with model uncertainty arising due to parametric error and unmodeled dynamics. This uncertainty can lead to unsafe or dangerous behavior, and thus it is critical to quantify safety in the face of uncertainty.

In this work we consider a data-driven, machine learning approach to reduce model uncertainty and improve the safe behavior of a system. Learning-based approaches have already shown great promise for controlling systems with uncertain models (\cite{schaal2010learning, kober2013reinforcement, khansari2014learning, chowdhary2015bayesian, cheng2019end, taylor2019episodic, shi2019neural, fan2019bayesian}). We look to achieve safety defined in terms of \textit{set invariance} (\cite{blanchini1999set, ames2019control}), which is an area of active research at the intersection of machine learning and control theory (\cite{berkenkamp2016safe, fisac2018general}). Additional details on related work are provided in Section \ref{sec:related}.

We will leverage Control Barrier Functions (CBFs) for synthesizing safe controllers (\cite{ames2014control, ames2017control}). CBFs have become increasingly popular for enforcing safety (\cite{nguyen2016exponential, wang2018safe}), but their safety guarantees require an accurate model of the system, and thus model uncertainty can lead to degradation of these guarantees (\cite{kolathaya2018input}). Robust CBF formulations can ensure safety in the face of model uncertainty (\cite{gurriet2018towards, xu2015robustness}), but may be overly restrictive of the system's behavior. Adaptive CBF methods can provide safety with arbitrarily large parametric error (\cite{taylor2019adaptive}), but only over a restricted class of model uncertainty.   

One challenge in utilizing learning for improving controllers is collecting training data that accurately reflects the desired learning environment. Exhaustive data collection suffers from the \textit{curse of dimensionality} and scales poorly with the size of the system's state and input. Additionally, the pre-collection of data upfront can lead to bad performance as the learning-augmented controller may enter new states not present in the pre-collected training data. We will utilize episodic learning approaches such as Datset Aggregation (DAgger) (\cite{Ross2010DAgger}) to address these challenges efficiently, and lead to an iteratively improved controller. Furthermore, we build upon recent work utilizing learning in the context of Control Lyapunov Functions (CLFs) (\cite{taylor2019episodic}) to construct an approach for learning model uncertainty. Unlike this preceding work, our method learns model uncertainty impacting safety rather than stability, and provides experimental demonstration. 

\textbf{Our contribution:} The contribution of this work is a novel approach for learning model uncertainty directly as it impacts the evolution of a Control Barrier Function and synthesizing a new controller for enforcing safety. Previous work has focused on learning an invariant set under a given controller, or learning unmodeled dynamics in the full-order state dynamics and modifying a controller. This approach learns the effect of unmodeled dynamics and parametric error on the system's behavior episodically, and integrates with optimization based methods for enforcing safety from nonlinear control theory. We demonstrate the functionality of this learning approach for a Segway system both in simulation and, for the first time, experimentally on a hardware platform.

Our paper is organized as follows. Section \ref{sec:CBFs} provides a review of Control Barrier Functions and safety-critical control. In Section \ref{sec:uncertain} we state assumptions on our model uncertainty, and define our learning approach for capturing the impact of this uncertainty on the CBF dynamics. Next, Section \ref{sec:episodic} details issues with standard supervised learning in a sequential prediction task, and provides an episodic framework that utilizes our particular learning approach. Lastly, Section \ref{sec:results} provides both simulation and experimental results on a Segway platform.

\section{Control Barrier Functions}
\label{sec:CBFs}
In this section we provide a review of Control Barrier Functions (CBFs) and how they can be utilized to ensure the safety of a system. This will motivate a particular episodic learning approach that achieves safety in the presence of uncertainty in modeled system dynamics.

Consider a dynamic system with state $\mb{x}\in\R^n$ and control input $\mb{u}\in\R^m$. We assume that the dynamics of the system are continuous in time and affine in the control input, such that:
\begin{equation}
\label{eqn:eom}
\dot{\mb{x}} = \mb{f}(\mb{x}) + \mb{g}(\mb{x})\mb{u} 
\qquad 
\xrightarrow{\mb{u} = \mb{k}(\mb{x})}
\qquad 
    \dot{\mb{x}} = \mb{f}_{\mathrm{cl}}(\mb{x}) = \mb{f}(\mb{x}) + \mb{g}(\mb{x})\mb{k}(\mb{x})
\end{equation}
where  $\mb{k}:\R^n\to\R^m$ is a locally Lipschitz continuous controller that yields closed-loop dynamics.   
Here $\mb{f}:\R^n \rightarrow \R^n$, $\mb{g}:\R^n \rightarrow \R^{n\times m}$ are assumed to be locally Lipschitz continuous and, therefore, $\mb{f}_{\mathrm{cl}}:\R^n\to\R^n$ is locally Lipschitz continuous.
Under this assumption, for any initial state $\mb{x}_0 \in \R^n$, there exists a time interval of existence, $I(\mb{x}_0) = [0, \tau_{max})$, such that there is a unique solution, $\mb{x}:I(\mb{x}_0)\to\R^n$, satisfying \eqref{eqn:eom} with $\mb{x}(0)=\mb{x}_0$ (\cite{perko2013differential}).

The notion of safety is formalized by specifying a \textit{safe set} that the state of the system must remain in to be considered safe. In particular, consider a set $\cal{S} \subset \R^n$ defined as the 0-superlevel set of a continuously differentiable function $h:\R^n \to \R$, with $0$ a regular value of $h$, yielding:
\begin{align}
    \cal{S} &\triangleq \{\mb{x} \in \R^n ~\vert~ h(\mb{x}) \geq 0\}. \label{eqn:S}
\end{align}
We refer to $\cal{S}$ as the \textit{safe set}, and note that $\partial\cal{S} \triangleq \{\mb{x} \in \R^n  ~\vert~ h(\mb{x}) = 0\}$ and $\interior(\cal{S}) \triangleq \{\mb{x} \in \R^n  ~\vert~ h(\mb{x}) > 0\}$. This motivates the following definitions of forward invariance and safety:
\begin{definition}[\textit{Invariance \& Safety}]
A set $\cal{S}$ is \textit{forward invariant} if for every $\mb{x}_0\in\cal{S}$, $\mb{x}(t) \in \cal{S}$ for all $t \in I(\mb{x}_0)$.  The system \eqref{eqn:eom} is \textit{safe} with respect to the set $\cal{S}$ if the set $\cal{S}$ is forward invariant.
\end{definition}
This definition implies that a system remains safe under a particular control policy only if all solutions to \eqref{eqn:eom} that begin in the safe set remain in the safe set. 

Control Barrier Functions serve as a synthesis tools for attaining the forward invariance, and thus safety, of a set (\cite{ames2014control,xu2015robustness}). Before defining CBFs, we note that a continuous function $\alpha: (-b,a) \to \R$ is an \textit{extended class} $\cal{K}$ function ($\alpha\in\cal{K}_e$) if $\alpha(0) = 0$ and $\alpha$ is strictly monotonically increasing. If $a,b = \infty$ and $\lim_{r\to \infty}\alpha(r) = \infty$ and $\lim_{r\to -\infty}\alpha(r) = -\infty$, then $\alpha$ is said to be an \textit{extended class} $\cal{K}_{\infty}$ function ($\alpha \in \cal{K}_{\infty,e})$. This enables the following definition of Control Barrier Functions:

\begin{definition}[\textit{Control Barrier Function (CBF), \cite{ames2017control}}]
Let $\cal{S} \subset \R^n$ be the 0-superlevel set of a continuously differentiable function $h:\R^n \rightarrow \R$ with $0$ a regular value. $h$ is a \textit{control barrier function} (CBF) for \eqref{eqn:eom} on $\cal{S}$ if there exists $\alpha\in\cal{K}_{\infty,e}$ such that for all $\mb{x}\in \cal{S}$:
\begin{align}
\label{eqn:CBF}
\sup_{\mb{u}\in \R^m}~ \dot{h}(\mb{x},\mb{u}) = 
    \sup_{\mb{u}\in \R^m}~\left[\derp{h}{\mb{x}}(\mb{x})\left(\mb{f}(\mb{x}) + \mb{g}(\mb{x})\mb{u}\right)\right] \geq -\alpha(h(\mb{x})).
\end{align}
\end{definition}
Given a CBF $h$ for \eqref{eqn:eom} and a corresponding $\alpha\in\cal{K}_{\infty,e}$, we can consider the point-wise set of all control values that satisfy \eqref{eqn:CBF}:
\begin{align*}
    K_{\mathrm{cbf}}(\mb{x}) = \left\{\mb{u}\in \R^m ~ \left| ~ \derp{h}{\mb{x}}(\mb{x})\left(\mb{f}(\mb{x}) + \mb{g}(\mb{x})\mb{u}\right) \geq -\alpha(h(\mb{x})) 
    \right. \right\},
\end{align*}
One of the main results of (\cite{ames2014control, xu2015robustness}) is that the existence of a CBF for $\cal{S}$ implies the system \eqref{eqn:eom} can be rendered safe with respect to $\cal{S}$:

\begin{theorem}
Given a set $\cal{S} \subset \R^n$ defined as the 0-superlevel set of a continuously differentiable function $h:\R^n\to\R$, if $h$ is a CBF for \eqref{eqn:eom} on $\cal{S}$, then any Lipshitz continuous controller $\mb{k}:\R^n\to\R^m$, such that $\mb{k}(\mb{x})\in K_{\mathrm{cbf}}(\mb{x})$ for all $\mb{x}\in\cal{S}$, renders the system \eqref{eqn:eom} safe with respect to the set $\cal{S}$.
\end{theorem}

This result motivates the construction of a point-wise optimal controller seeking to minimize a cost associated with the choice of input. To this end, we consider the \textit{safety-critical control} formulation (\cite{gurriet2018towards}) that seeks to filter a hand-designed but potentially unsafe Lipschitz continuous controller, $\mb{k}_d:\R^n\to\R^m$, to find the nearest safe action:
\begin{align}
\label{eqn:CBF-QP}
\tag{CBF-QP}
\mb{k}(\mb{x}) =  \,\,\underset{\mb{u} \in \R^m}{\argmin}  &  \quad \frac{1}{2} \| \mb{u}-\mb{k}_d(\mb{x}) \|_2^2  \\
\mathrm{s.t.} \quad & \quad \frac{\partial h}{\partial \mb{x}}(\mb{x})\left(\mb{f}(\mb{x}) +\mb{g}(\mb{x})\mb{u} \right)
\geq -\alpha(h(\mb{x})) \nonumber
\end{align}
The validity of $h$ as a CBF ensures the feasibility of this optimization problem, and the resulting controller is Lipschitz continuous (\cite{morris2013sufficient, jankovic2018robust}). As the dynamics model \eqref{eqn:eom} appears in the constraint enforcing safety, the guarantees of safe control synthesis endowed by this controller require strong assumptions on the accuracy of the dynamics model.

\section{Uncertainty Models \& Learning}
\label{sec:uncertain}
This section provides structural assumptions on the model uncertainty present in our system, and defines our dynamics learning problem in the context of safety. In practice, uncertainty in the dynamics of the system exists due to parametric error and unmodeled dynamics, such that the functions $\mb{f}$ and $\mb{g}$ are not precisely known. Rather, control synthesis is done with a nominal model that estimates the true dynamics of the system:
\begin{equation}
\label{eqn:nom_eom}
    \widehat{\dot{\mb{x}}} = \widehat{\mb{f}}(\mb{x}) + \widehat{\mb{g}}(\mb{x})\mb{u}
\end{equation}
where $\widehat{\mb{f}}:\R^n\to\R^n$ and $\widehat{\mb{g}}:\R^n\to\R^{n\times m}$ are Lipschitz continuous. By adding and subtracting this expression to \eqref{eqn:eom}, we see that the system evolves under the differential equation given by:
\begin{equation}
\label{eqn:res_eom}
    \dot{\mb{x}} = \widehat{\mb{f}}(\mb{x}) + \widehat{\mb{g}}(\mb{x})\mb{u} + \underbrace{\mb{f}(\mb{x})-\widehat{\mb{f}}(\mb{x})}_{\mb{b}(\mb{x})} + \underbrace{\left(\mb{g}(\mb{x})-\widehat{\mb{g}}(\mb{x})\right)}_{\mb{A}(\mb{x})}\mb{u}
\end{equation}
where $\mb{b}:\R^n\to\R^n$ and $\mb{A}:\R^n\to\R^{n\times m}$ are the unmodeled dynamics. This uncertainty in the dynamics additionally manifests in the time of derivative of CBFs designed for the system:
\begin{equation}
\label{eqn:res_h}
    \dot{h}(\mb{x},\mb{u}) = \underbrace{\derp{h}{\mb{x}}(\mb{x})(\widehat{\mb{f}}(\mb{x})+\widehat{\mb{g}}(\mb{x})\mb{u})}_{\widehat{\dot{h}}(\mb{x},\mb{u})} + \underbrace{\derp{h}{\mb{x}}(\mb{x})\mb{b}(\mb{x})}_{b(\mb{x})} + \underbrace{\derp{h}{\mb{x}}(\mb{x})\mb{A}(\mb{x})}_{\mb{a}(\mb{x})^\top}\mb{u}
\end{equation}
where $b:\R^n\to\R$ and $\mb{a}:\R^n\to\R^m$. The presence of uncertainty in the CBF time derivative makes it impossible to verify if a given control input satisfies the inequality in \eqref{eqn:CBF}, and can lead to unsafe behavior. One critical assumption we make on the uncertainty in the dynamics is as follows:

\begin{assumption}
If a function $h$ is a valid CBF for the nominal model of the system \eqref{eqn:nom_eom}, then it is a valid CBF for the uncertain dynamic system \eqref{eqn:res_eom}.
Mathematically this assumption appears as:
\begin{equation*}
    \sup_{\mb{u}\in\R^m}\derp{h}{\mb{x}}(\mb{x})\left(\widehat{\mb{f}}(\mb{x})+\widehat{\mb{g}}(\mb{x})\mb{u}\right)\geq-\alpha(h(\mb{x})) \implies \sup_{\mb{u}\in\R^m}\derp{h}{\mb{x}}(\mb{x})\left(\mb{f}(\mb{x})+\mb{g}(\mb{x})\mb{u}\right)\geq-\alpha(h(\mb{x}))
\end{equation*}
\end{assumption}
Intuitively this assumption requires that a set within the state space that can be keep safe for the nominal model of the system can also be kept safe for the uncertain system. Technically this assumption amounts to an assumption that the relative degree of $h$ for the model is the same as the relative degree of $h$ for the true system. This assumption on the true system naturally follows if the true system retains the same actuation capability as the model, with more technical details provided in \cite{Sastry99}. While this assumption may enable a robust CBF approach for certifying safety, this method may be overly conservative. Instead, we take a data-driven approach similar to (\cite{taylor2019episodic}) to learn uncertainty as it appears in the time derivative of the CBF, $\dot{h}$, given in \eqref{eqn:res_h}. 

To motivate our learning framework, consider a simple approach for learning $\mb{a}$ and $b$ via supervised regression (\cite{gyorfi2006distribution}): an experiment is conducted using a nominal controller to collect data and learn functions that approximate $\mb{a}$ and $b$ via supervised regression. More concretely, an experiment consists of sampling an initial state for the system from an initial state distribution $\cal{X}_0$, and allowing the system to evolve for a finite time interval under a given state-feedback controller (using a sample-hold implementation). This yields a discrete time state and input history, allowing the computation of a discrete time history of the CBF $h$. This time history is numerically differentiated to compute an approximate time history of the true value of the time derivative $\dot{h}$, yielding a data-set of input-output pairs:
\begin{equation}
    \mathfrak{D} = \{((\mb{x}_i, \mb{u}_i), \dot{h}_i)\}_{i=1}^N \subseteq (\R^n \times \R^m) \times \R.
\end{equation}
Consider two nonlinear function classes $\cal{H}_{\mb{a}}:\R^n\to\R^m$ and $\cal{H}_{b}:\R^n\to\R$. For $\widehat{\mb{a}}\in\cal{H}_a$ and $\widehat{b}\in\cal{H}_b$, we define the estimator:
\begin{equation}
    \label{eqn:estimator}
    \widehat{\dot{S}}(\mb{x},\mb{u}) = \widehat{\dot{h}}(\mb{x},\mb{u})+\widehat{\mb{a}}(\mb{x})^\top\mb{u} + \widehat{b}(\mb{x})
\end{equation}
and let $\cal{H}$ be the class of all such estimators mapping $\R^n\times\R^m$ to $\R$. Specifying a loss function $\cal{L}:\R\times\R\to\R$, the supervised regression task is to find an estimator in $\cal{H}$ via empirical risk minimization (ERM):
\begin{equation}
    \label{eqn:erm}
    \inf_{\substack{\widehat{\mb{a}}\in\cal{H}_a \\ \widehat{b}\in\cal{H}_b}}\frac{1}{N}\sum_{i=1}^N\cal{L}\left(\widehat{\dot{S}}(\mb{x}_i,\mb{u}_i),\dot{h}_i\right)
\end{equation}
This experimental framework can be utilized in both simulation (learning estimates of the CBF dynamics when model information is withheld from the state-feedback controller), or directly on hardware platforms (when a given model does not accurately describe the physical system). 

\section{Episodic Learning Framework}
\label{sec:episodic}
In this section we discuss the consequences of using conventional supervised learning in this context, and formulate an episodic learning framework to overcome these consequences. A critical assumption of supervised learning is independently and identically distributed (i.i.d) training data. The data collected via the preceding experimental framework violates this assumption, as each data point (state and input pair) depends on the value of preceding data points. One potential consequence of violating this assumption when using standard supervising learning is error cascades (\cite{le2016smooth}). To overcome the impacts of violating this assumption, we extend this experimental protocol to an episodic learning framework.

In particular, this framework iteratively alternates between running experiments with an intermediate controller (or roll-outs in reinforcement learning (\cite{kober2013reinforcement})) to collect data, and using the newly collected data to synthesize a new controller. Thus learning $\mb{a}$ and $b$ is integrated with improving the safety of the system. The new controller can be synthesized similarly to the safety-critical controller \eqref{eqn:CBF-QP}, but utilizing the updated estimate of the CBF time derivative given in \eqref{eqn:estimator}:
\begin{align}
\label{eqn:LCBF-QP}
\tag{LCBF-QP}
\mb{k}'(\mb{x}) =  \,\,\underset{\mb{u} \in \R^m}{\argmin}  &  \quad \frac{1}{2} \| \mb{u}-\mb{k}_d(\mb{x}) \|_2^2  \\
\mathrm{s.t.} \quad & \quad \widehat{\dot{S}}(\mb{x},\mb{u})
\geq -\alpha(h(\mb{x})) \nonumber
\end{align}
This controller effectively finds the nearest control input to the hand-designed controller $\mb{k}_d$ that renders the system safe according to the estimator \eqref{eqn:estimator}. Thus, the actual safety achieved by the system degrades if there is error present in \eqref{eqn:estimator}.

To mitigate this remaining error, we use the newly synthesized controller to improve our estimator. One method for reducing this error is to run experiments and use conventional supervised regression to update the estimator. As previously noted, conventional supervised regression faces limitations with non-i.i.d data. These limitations can be overcome episodically by utilizing reduction techniques, whereby a sequential prediction problem is reduced to a sequence of supervised learning problems (\cite{ernst2005tree, ross2011reduction}). In each episode, data is generated using a different controller. The data set is aggregated with all previously collected data and a new ERM problem \eqref{eqn:erm} is solved over the full data set. Our approach draws inspiration from the DAgger algorithm (\cite{Ross2010DAgger}), with a key difference: DAgger is a model-free policy learning algorithm, which trains a policy directly with an optimal policy oracle. Our method does not receive information on the optimal policy nor does it train a policy directly, but learns information about the unmodeled dynamics impacting safety and indirectly determines a policy via \eqref{eqn:LCBF-QP}. 

Lastly, as the newly synthesized controller will expose the system to new regions of the state space in which the estimator may not be accurate, the synthesized controller is blended with the nominal controller using heuristically chosen weights, $w_j\in[0,1]$, that reflect the trust in the learned model. This effectively limits how quickly the system's behavior can change. The episodic framework is summarized in Algorithm \ref{alg:dacbarf}.

\begin{algorithm}[h]
    \SetKwData{Left}{left}
    \SetKwFunction{Sample}{sample}\SetKwFunction{Experiment}{experiment} \SetKwFunction{ERM}{ERM}\SetKwFunction{Augment}{augment}
    \SetKwInOut{Input}{input} \SetKwInOut{Output}{output}
    
    \Input{CBF $h$, CBF derivative estimate $\widehat{\dot{h}}_0$, model classes $\cal{H}_{\mb{a}}$ and $\cal{H}_b$, loss function $\cal{L}$, set of initial conditions $\cal{X}_0$, nominal state-feedback controller $\mb{k}_0$, number of experiments $T$, sequence of trust coefficients $0 \leq w_1 \leq \cdots \leq w_T \leq 1$}
    \Output{Dataset $D$, CBF derivative estimate $\widehat{\dot{h}}_T$, augmented controller $\mb{k}_T$}
    \BlankLine
    $D = \emptyset$\; \tcp*[f]{Initialize data set}\;\\
    \For{$j = 1,\ldots,T$}{
    $\mb{x}_0 \leftarrow$\Sample{$\cal{X}_0$} \tcp*[f]{Sample initial condition}\;\\
    $D_j \leftarrow$\Experiment{$\mb{x}_0,\mb{k}_{j-1}$} \tcp*[f]{Execute experiment}\;\\
    $D\leftarrow D\cup D_j$ \tcp*[f]{Aggregate dataset}\;\\
    $\widehat{\mb{a}},\widehat{b}\leftarrow$\ERM{$\cal{H}_{\mb{a}},\cal{H}_b,\cal{L},D,\widehat{\dot{h}}_0$} \tcp*[f]{Fit estimators}\;\\
    $\widehat{\dot{h}}_j\leftarrow\widehat{\dot{h}}_0+\widehat{\mb{a}}^\top\mb{u}+\widehat{b}$ \tcp*[f]{Update derivative estimator}\;\\
    $\mb{k}_j\leftarrow(1-w_j)\cdot\mb{k}_0+w_j~\cdot~$\Augment{$\widehat{\dot{h}}_j$} \tcp*[f]{Update controller}\;
    }
\caption{Dataset Aggregation for Control Barrier Functions (DaCBarF)}
\label{alg:dacbarf}
\end{algorithm}

\section{Simulation \& Experimental Results}
\label{sec:results}
\begin{figure*}[h]
     \centering
     \begin{subfigure}
        {\includegraphics[clip ,scale =0.13, valign =t ]{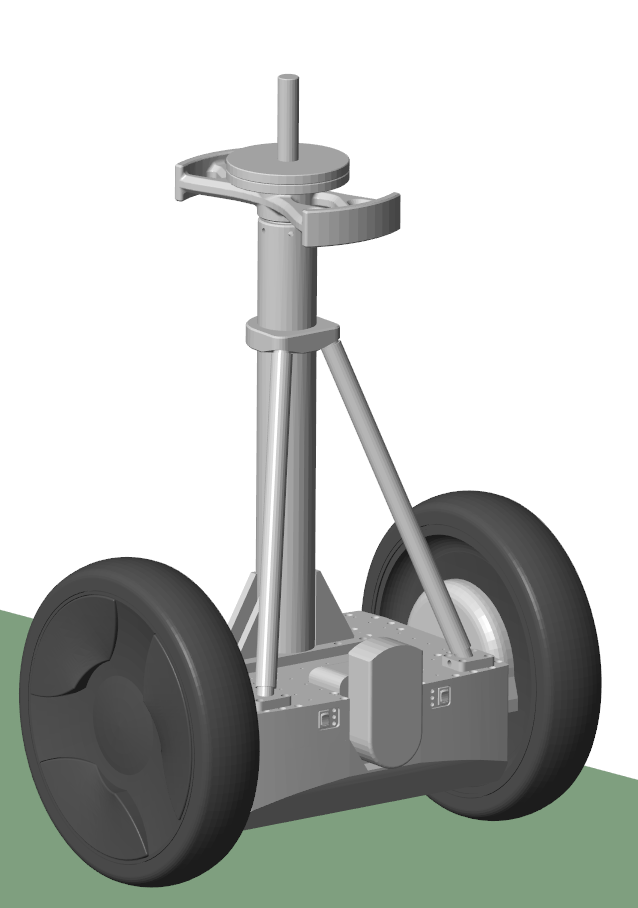}}
    \end{subfigure}
    \begin{subfigure}
        {\includegraphics[clip, scale = 0.4, valign = t]{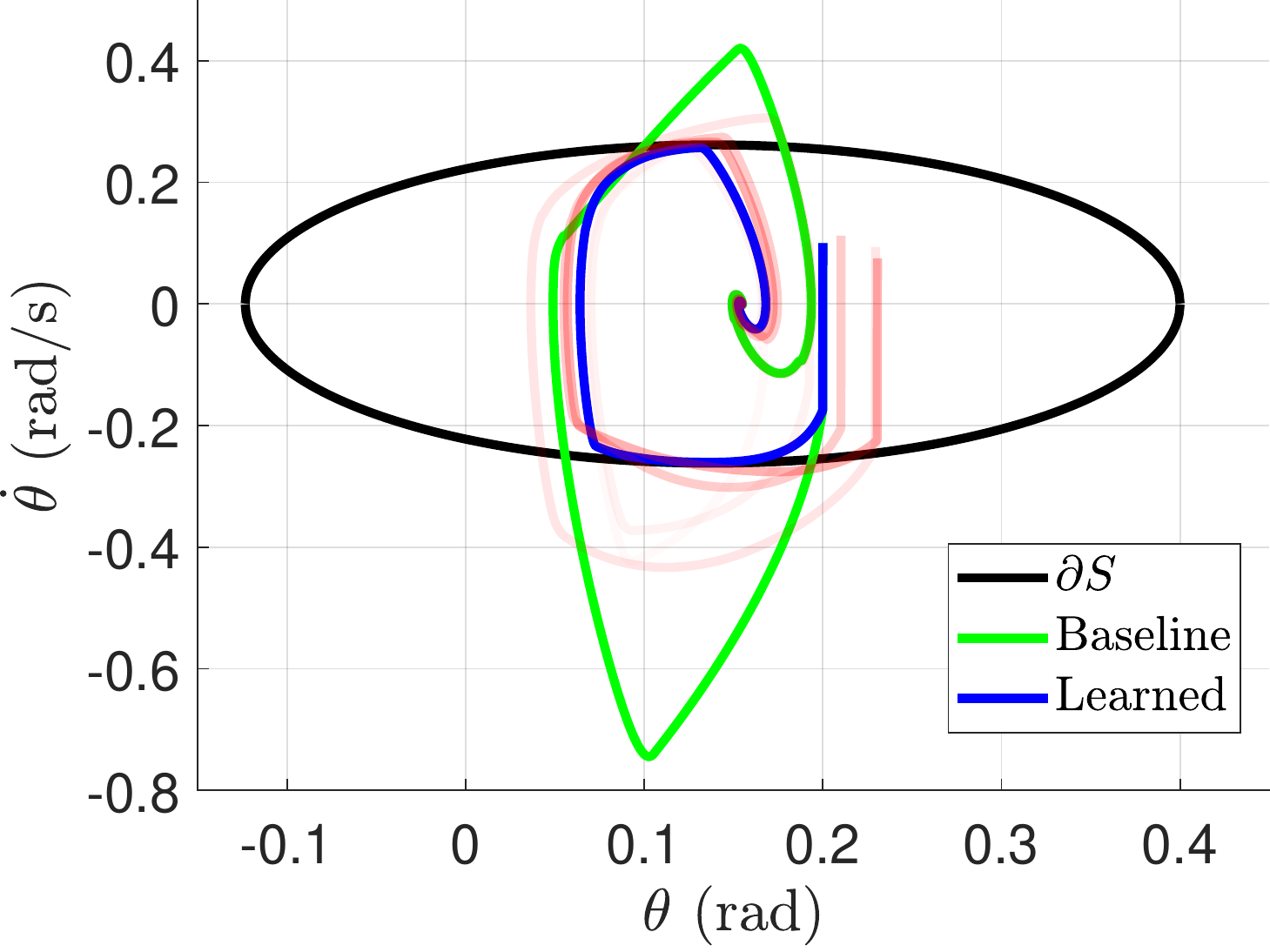}}
    \end{subfigure}
     \begin{subfigure}
        {\includegraphics[clip, scale = 0.4, valign =  t]{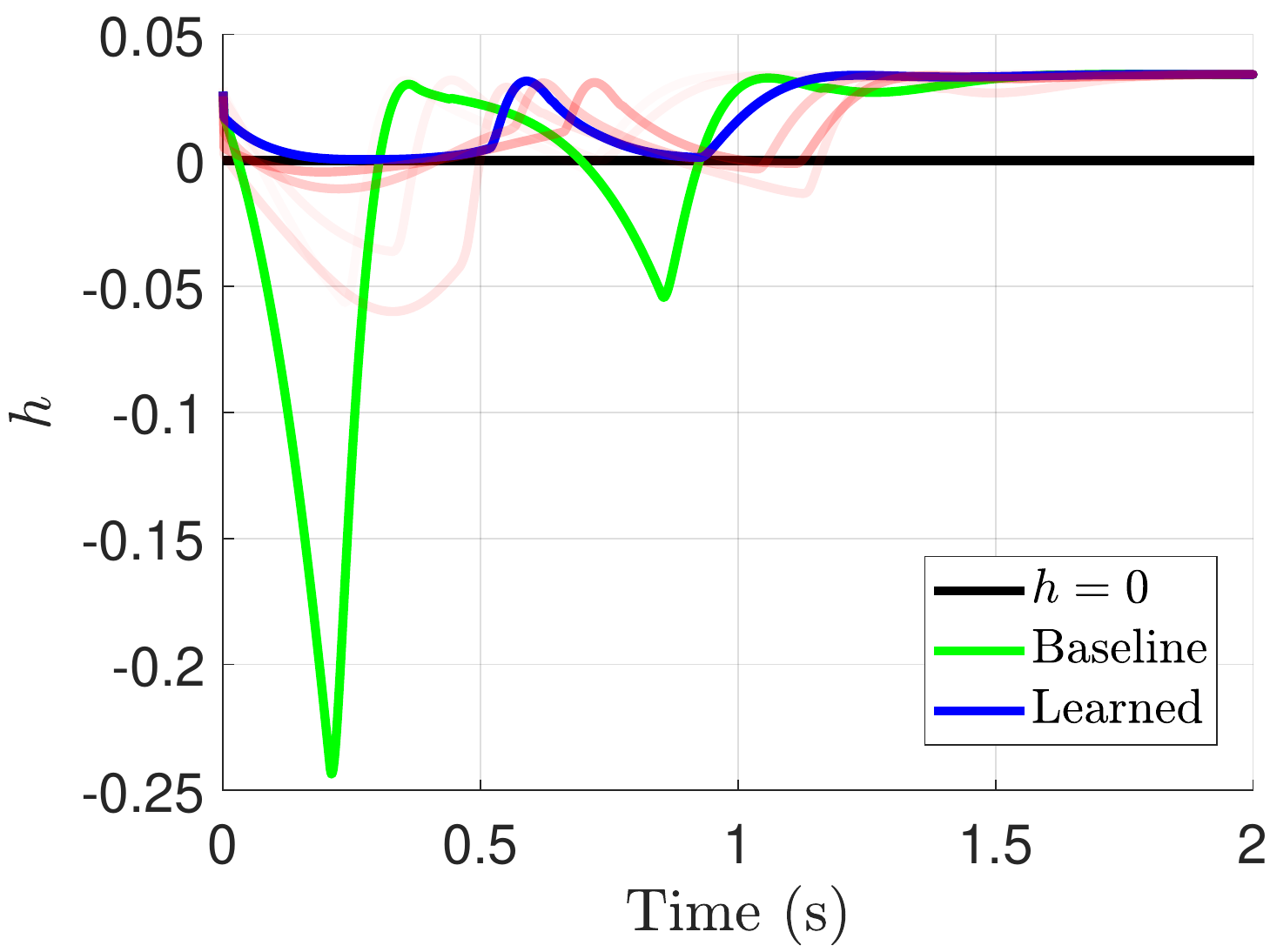}}
    \end{subfigure}
    \caption{Simulation results using the model-based CBF controller \eqref{eqn:CBF-QP} (green) and learning-augmented controller \eqref{eqn:LCBF-QP} (blue). Episodic data appears as red traces. (a) Segway dynamics simulation CAD model. (b) The phase portrait showing the evolution of the state of the system leaves the safe set (black ellipse) with the model-based controller, while it remains within using the learning-augmented controller. (c) The value of $h$ drops below zero (twice) using the model-based controller, while it remains above zero using the learning-augmented controller. Video can be found at \url{https://vimeo.com/user106627792/review/380798349/649d4c391e}} 
    \label{fig:sim_results}
\end{figure*}

\begin{figure*}[h]
    \centering
        \begin{subfigure}
        {\includegraphics[trim={8.5in, .1in, 7in, 1in}, clip, scale =0.1, valign =t ]{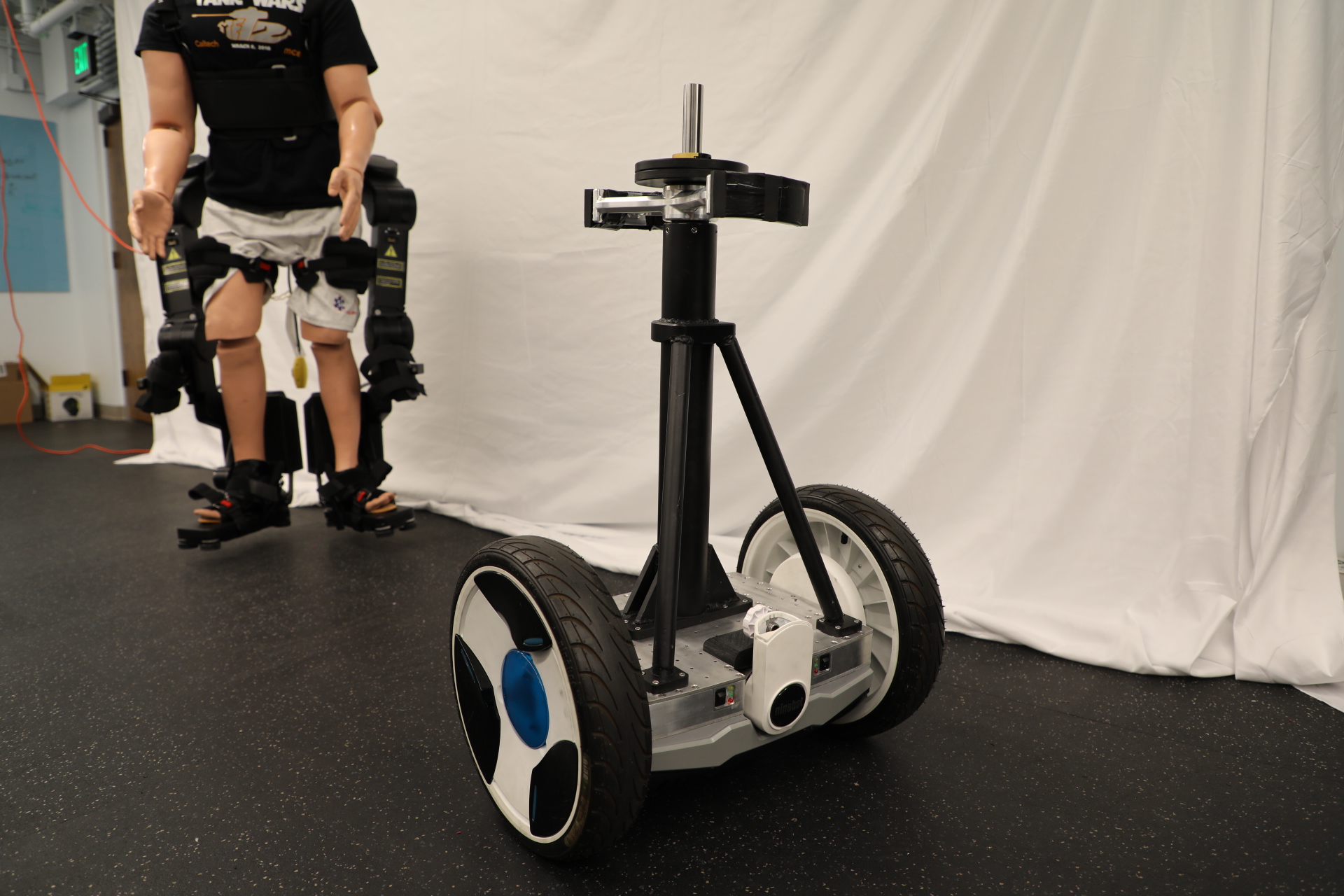}}
    \end{subfigure}
        \begin{subfigure}
        {\includegraphics[clip, scale = 0.4, valign = t]{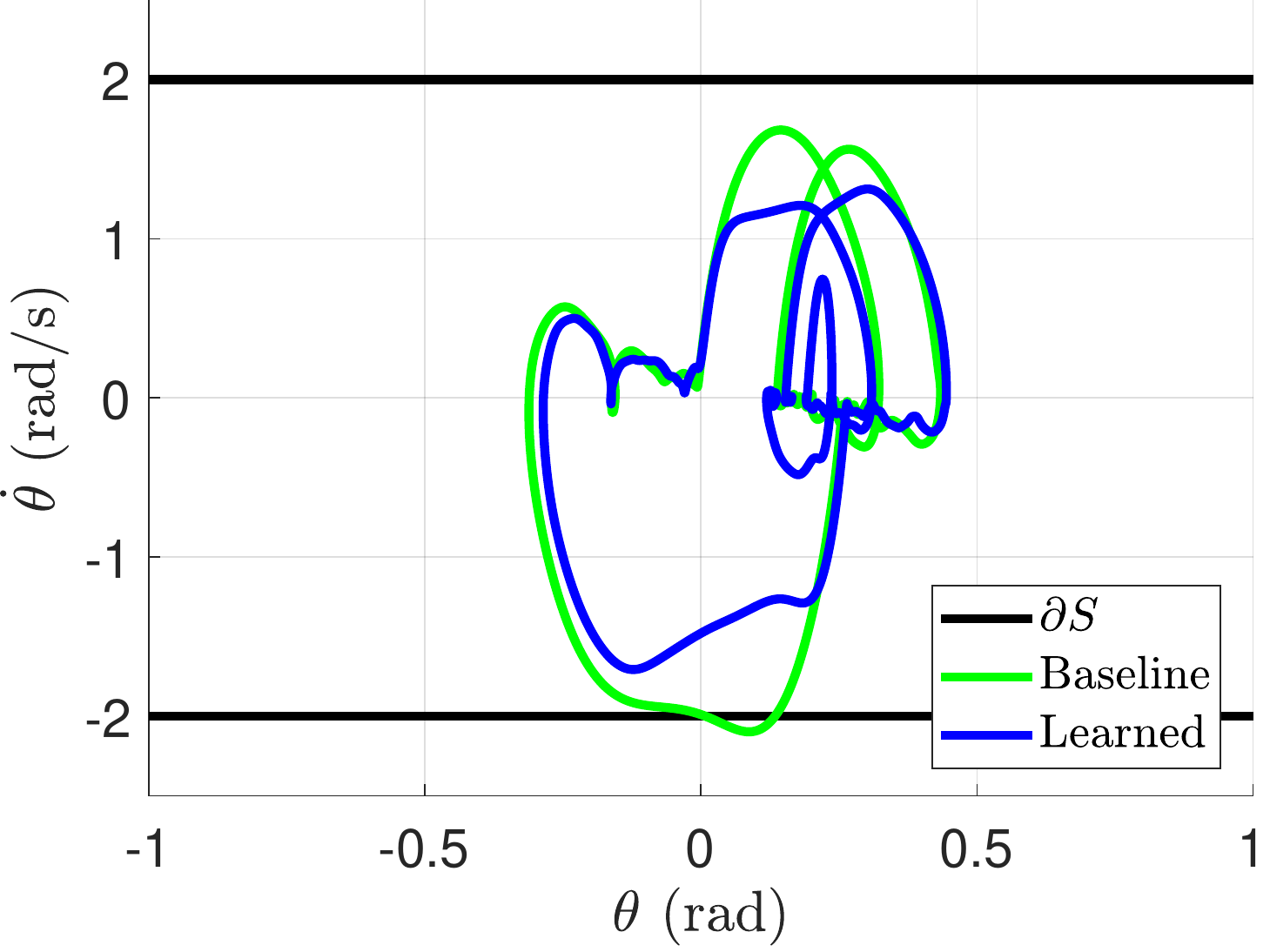}}
    \end{subfigure}
     \begin{subfigure}
        {\includegraphics[clip, scale = 0.4, valign =  t]{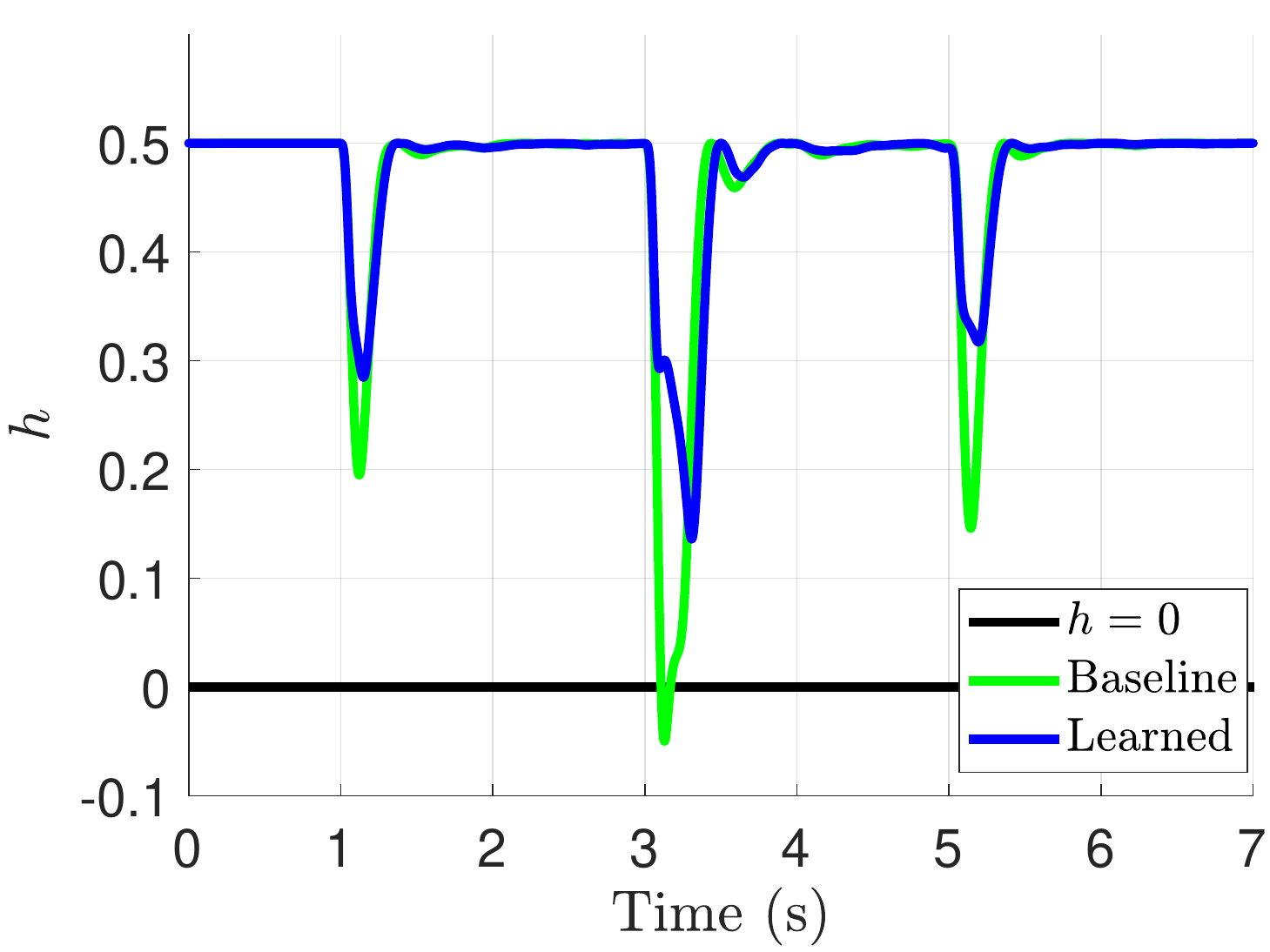}}
    \end{subfigure}
    \caption{Experimental results using the model-based CBF controller \eqref{eqn:CBF-QP} (green) and learning-augmented controller \eqref{eqn:LCBF-QP} (blue). (a) Custom robotic Ninebot Segway platform. (b) The phase portrait showing the evolution of the state of the system leaves the safe set (black strip) with the model-based controller, while it remains within using the learning-augmented controller. (c) The value of $h$ drops below zero (at $t=3$) using the model-based controller, while it remains above zero using the learning-augmented controller. Video can be found at \url{https://vimeo.com/user106627792/review/380798276/f37b003db3}}
    \label{fig:exp_results}
\end{figure*}

In this section we apply the episodic learning algorithm developed in Sections \ref{sec:uncertain} and \ref{sec:episodic} to the Segway platform. In particular, we consider a 4-dimensional asymmetrical planar Segway model as in \cite{gurriet2018towards}, seen in Figure \ref{fig:sim_results} and \ref{fig:exp_results}. The state of the system consists of the horizontal position, $x$, velocity, $\dot{x}$, pitch angle, $\theta$, and pitch angle rate $\dot{\theta}$. Control input to a motor in the system's wheel is specified as a percentage of the systems battery voltage, which leads to a torque. This model permits an affine control system as per \eqref{eqn:eom}. Stabilization of this system to an equilibrium configuration is done using a hand-tuned proportional-derivative (PD) controller, $\mb{k}_d$.

Safety of the Segway system is often encoded as limitations on how far and how fast the Segway is allowed to tip from an equilibrium configuration. This appears mathematically as the Control Barrier Function on the pitch angle and pitch angle rate:
\begin{equation}
    \label{eqn:segwaycbf}
    h(\theta,\dot{\theta}) = \frac{1}{2}\left(\theta_{max}^2 - (\theta-\theta_e)^2-c\dot{\theta}^2\right),
\end{equation}
where $\theta_{max}$ is a limit on how far the Segway can tip, $\theta_e$ is an equilibrium angle, and the coefficient $c$ scales the importance of $\dot{\theta}$ in defining safety. As $h$ is a CBF, the controller \eqref{eqn:CBF-QP} can be used to ensure the safety of the system.

To evaluate the impact of model uncertainty on the CBF's ability to ensure the safety of the system, we perturb the mass and electrical parameters of the Segway model by up to 15\%, but withhold this uncertain model from the controller \eqref{eqn:CBF-QP}. Figure \ref{fig:sim_results} shows that the model-based controller is unable to achieve safety, with the state of the system leaving the 0-superlevel set of $h$ (left), and the value of $h$ falling below $0$ (right). 

We deploy our episodic learning framework on the uncertain Segway system to overcome this model uncertainty and achieve safety. In particular, we conduct a sequence of 10 experiments with varying initial conditions reflecting disturbances to the Segway. Data from each experiment is aggregated into one dataset that is used to train regression models. For this system we use neural-networks with one hidden layer of 200 nodes for the estimators $\widehat{\mb{a}}$ and $\widehat{b}$. These networks are trained to minimize mean absolute error using stochastic gradient descent (SGD)\footnote{The models were implemented and trained in Python using the Keras learning package.}. After each episode the controller \eqref{eqn:LCBF-QP} is blended with the original controller \eqref{eqn:CBF-QP} using weights that grow linearly across the episodes. Figure \ref{fig:sim_results} shows the behavior of the system when the controller \eqref{eqn:LCBF-QP} is deployed after episodic learning. We see that the state of the system is kept within the 0-superlevel set of $h$ (left), with the value of $h$ being kept above $0$ (right). This indicates that the learning-augmented controller is able to keep the system safe even with model uncertainty.

To further demonstrate the ability of this learning approach, we deployed it on the physical Segway system seen in Figure \ref{fig:exp_results}. The objective of the Segway was specified as tracking a desired horizontal velocity profile while satisfying a barrier function on the pitch angle rate specified as:
\begin{equation}
h(\dot{\theta}) = \frac{1}{2}(1-c\dot{\theta}^2)
\end{equation}
The desired velocity profile led to a violation of the barrier function when only enforcing the safety constraint using model information. A series of 3 episodes were conducted tracking this velocity profile without introducing learned information. The estimators $\widehat{\mb{a}}$ and $\widehat{b}$ were implemented as neural networks with two hidden layers of 50 nodes and trained on this data after which they were incorporated into the controller using $w_j=1$. This high weighting was used to see noticeable changes in behavior of the Segway after learning. The improvement in safety using learning can be seen in Figure \ref{fig:exp_results}.

\section{Related Work}
\label{sec:related}
Our work combines the idea of Control Barrier Functions from model-based control theory with elements from supervised learning, developing an episodic framework for directly learning model uncertainty as it impacts safe behavior. This relates to the idea of using model-based methods for ensuring safety with learning to reduce model uncertainty more broadly, as detailed below.

\textbf{Forward Invariance with Synthesis:}
Learning for forward invariance with synthesis focuses on learning how to choose control inputs that renders a desired region forward invariant. \cite{cheng2019end} deploys reinforcement learning with barrier functions to develop an estimate of the dynamics and an optimal policy. In \cite{saveriano2019learning}, human demonstrations with robotic arms are used as motion primitives to learn barrier functions that describe safe regions. \cite{wang2018safe} uses \textit{Gaussian Processes (GPs)} in conjunction with barrier functions to explore a region which is certifiably forward invariant in the context of quadrotors. Using methods from adaptive control, \cite{taylor2019adaptive, fan2019bayesian} update estimates of CBF dynamics to ensure safety. Without using barrier functions, \cite{aswani2013provably, koller2018learning} deploy learning in the context of \textit{model-predictive control (MPC)}, which choose control inputs via dynamic programming to avoid unsafe states.

\textbf{Region-of-Attraction Estimation:} Region-of-Attraction (ROS) estimation focuses on safely learning a region in the state space from which the system is guaranteed to converge to a desired equilibrium point. Much of this theory is motivated by the notion of Lyapunov stability from nonlinear control theory (\cite{Khalil}), which implies a stricter definition of safety as it requires not only forward invariance, but also convergence. \cite{berkenkamp2016safe2} begins with a stabilizing controller and safely explores the ROA for that controller. \cite{berkenkamp2017safe} develops upon this using ROA in conjunction with reinforcement learning to safely to improve a policy to maximize the ROA. \cite{taylor2019control} develops a data-driven approach to evaluate regions over which \textit{projection-to-state stability} (\cite{sontag1989universal}) holds, studying how learning error impacts stability.

\textbf{Learning Lyapunov Functions:} Learning Lyapunov functions encompasses a variety of different approaches for certifying  the stricter condition of stability to ensure safety guarantees. \cite{chow2018lyapunov} focuses on learning Lyapunov functions for discrete Markov Decision Processes. \cite{richards2018lyapunov} uses a neural-network to model a Lyapunov function that describes a maximum ROA, and \cite{gallieri2019safe} learns both a discrete Lyapunov function and a stabilizing policy. \cite{khansari2014learning} focuses on learning Control Lyapunov Functions (CLFs) (\cite{artstein1983stabilization}) from data for robotic reaching motions, while \cite{taylor2019episodic} motivates the work in this paper in the context of CLFs by learning uncertainty in a given CLF's time derivative. While explicitly not learning Lyapunov functions, \cite{berkenkamp2015safe} ensures stability and thus safety during learning for linear systems.

\textbf{Reachable Set Learning:} Reachable Set learning focuses on learning what states are reachable by the system and developing policies to limit the reachable set so as to avoid unsafe states. \cite{gillula2011guaranteed,gillula2012guaranteed} formulates an approach to safe online learning via \textit{Hamilton-Jacobi-Isaacs (HJI)} reachability and demonstrates it in a target-tracking context. \cite{akametalu2014reachability} uses GPs to estimate the unmodeled dynamics and better quantify the reachable set. \cite{fisac2018general} summarizes these preceding results theoretically and demonstrates it experimentally.

\textbf{Performance-Based Safety:} Performance-based safety relates safety to a minimum level of performance achieved by a deployed policy. \cite{sui2015safe} develops an algorithmic approach for safely exploring different policies while maintaining a performance threshold. The work in \cite{berkenkamp2016safe} extends this algorithmic work to tuning controller parameters for discrete dynamic systems and demonstrates it on a quadrotor platform.

\section{Conclusion}
We presented an episodic learning framework that integrates with Control Barrier Functions, a model-based nonlinear control method for ensuring safety. Our method is able to learn unmodeled dynamics and parametric error as they directly impact the safety of the system and incorporate this information into an optimization based controller. We demonstrate the ability of this method to improve the safety of a Segway system in simulation and experimentally on hardware.


\bibliography{main}

\end{document}